\documentclass[twocolumn]{aastex631}

\graphicspath{{./}{figures/}}

\begin{document}

\title{Minute-cadence observations on Galactic plane with Wide Field Survey Telescope (WFST):\\
Overview, methodology and early results
}

\author[0000-0003-3965-6931]{Jie Lin}
\affiliation{Department of Astronomy, University of Science and Technology of China, Hefei 230026, People's Republic of China}
\altaffiliation{linjie2019@ustc.edu.cn}
\affiliation{School of Astronomy and Space Science, University of Science and Technology of China, Hefei 230026, People's Republic of China}

\author{Tinggui Wang}
\affiliation{Department of Astronomy, University of Science and Technology of China, Hefei 230026, People's Republic of China}
\affiliation{School of Astronomy and Space Science, University of Science and Technology of China, Hefei 230026, People's Republic of China}
\affiliation{Deep Space Exploration Laboratory, Hefei 230088, People's Republic of China}

\author{Minxuan Cai}
\affiliation{Department of Astronomy, University of Science and Technology of China, Hefei 230026, People's Republic of China}
\affiliation{School of Astronomy and Space Science, University of Science and Technology of China, Hefei 230026, People's Republic of China}

\author{Zhen Wan}
\affiliation{Department of Astronomy, University of Science and Technology of China, Hefei 230026, People's Republic of China}
\affiliation{School of Astronomy and Space Science, University of Science and Technology of China, Hefei 230026, People's Republic of China}

\author{Xuzhi Li}
\affiliation{School of Mathematics and Physics, Anqing Normal University, Anqing 246133, People's Republic of China}
\affiliation{Institute of Astronomy and Astrophysics, Anqing Normal University, Anqing 246133, People's Republic of China}

\author{Lulu Fan}
\affiliation{Department of Astronomy, University of Science and Technology of China, Hefei 230026, People's Republic of China}
\affiliation{School of Astronomy and Space Science, University of Science and Technology of China, Hefei 230026, People's Republic of China}
\affiliation{Deep Space Exploration Laboratory, Hefei 230088, People's Republic of China}

\author{Qingfeng Zhu}
\affiliation{Department of Astronomy, University of Science and Technology of China, Hefei 230026, People's Republic of China}
\affiliation{School of Astronomy and Space Science, University of Science and Technology of China, Hefei 230026, People's Republic of China}

\author{Ji-an Jiang}
\affiliation{Department of Astronomy, University of Science and Technology of China, Hefei 230026, People's Republic of China}
\affiliation{School of Astronomy and Space Science, University of Science and Technology of China, Hefei 230026, People's Republic of China}

\author{Ning Jiang}
\affiliation{Department of Astronomy, University of Science and Technology of China, Hefei 230026, People's Republic of China}
\affiliation{School of Astronomy and Space Science, University of Science and Technology of China, Hefei 230026, People's Republic of China}

\author{Xu Kong}
\affiliation{Department of Astronomy, University of Science and Technology of China, Hefei 230026, People's Republic of China}
\affiliation{School of Astronomy and Space Science, University of Science and Technology of China, Hefei 230026, People's Republic of China}
\affiliation{Deep Space Exploration Laboratory, Hefei 230088, People's Republic of China}

\author{Zheyu Lin}
\affiliation{Department of Astronomy, University of Science and Technology of China, Hefei 230026, People's Republic of China}
\affiliation{School of Astronomy and Space Science, University of Science and Technology of China, Hefei 230026, People's Republic of China}

\author{Jiazheng Zhu}
\affiliation{Department of Astronomy, University of Science and Technology of China, Hefei 230026, People's Republic of China}
\affiliation{School of Astronomy and Space Science, University of Science and Technology of China, Hefei 230026, People's Republic of China}

\author{Zhengyan Liu}
\affiliation{Department of Astronomy, University of Science and Technology of China, Hefei 230026, People's Republic of China}
\affiliation{School of Astronomy and Space Science, University of Science and Technology of China, Hefei 230026, People's Republic of China}

\author{Jie Gao}
\affiliation{State Key Laboratory of Particle Detection and Electronics, University of Science and Technology of China, Hefei 230026, People's Republic of China}

\author{Bin Li}
\affiliation{Purple Mountain Observatory, Chinese Academy of Sciences, Nanjing 210023, People's Republic of China}

\author{Feng Li}
\affiliation{State Key Laboratory of Particle Detection and Electronics, University of Science and Technology of China, Hefei 230026, People's Republic of China}

\author{Ming Liang}
\affiliation{National Optical Astronomy Observatory (NSF’s National Optical-Infrared Astronomy Research Laboratory), 950 N Cherry Ave., Tucson, AZ 85726, USA}

\author{Hao Liu}
\affiliation{State Key Laboratory of Particle Detection and Electronics, University of Science and Technology of China, Hefei 230026, People's Republic of China}

\author{Wei Liu}
\affiliation{Purple Mountain Observatory, Chinese Academy of Sciences, Nanjing 210023, People's Republic of China}

\author{Wentao Luo}
\affiliation{Deep Space Exploration Laboratory, Hefei 230088, People's Republic of China}

\author{Jinlong Tang}
\affiliation{Institute of Optics and Electronics, Chinese Academy of Sciences, Chengdu 610209, People's Republic of China}

\author{Hairen Wang}
\affiliation{Purple Mountain Observatory, Chinese Academy of Sciences, Nanjing 210023, People's Republic of China}

\author{Jian Wang}
\affiliation{State Key Laboratory of Particle Detection and Electronics, University of Science and Technology of China, Hefei 230026, People's Republic of China}
\affiliation{Deep Space Exploration Laboratory, Hefei 230088, People's Republic of China}

\author{Yongquan Xue}
\affiliation{Department of Astronomy, University of Science and Technology of China, Hefei 230026, People's Republic of China}
\affiliation{School of Astronomy and Space Science, University of Science and Technology of China, Hefei 230026, People's Republic of China}

\author{Dazhi Yao}
\affiliation{Purple Mountain Observatory, Chinese Academy of Sciences, Nanjing 210023, People's Republic of China}

\author{Hongfei Zhang}
\affiliation{State Key Laboratory of Particle Detection and Electronics, University of Science and Technology of China, Hefei 230026, People's Republic of China}

\author{Xiaoling Zhang}
\affiliation{Purple Mountain Observatory, Chinese Academy of Sciences, Nanjing 210023, People's Republic of China}

\author{Wen Zhao}
\affiliation{Department of Astronomy, University of Science and Technology of China, Hefei 230026, People's Republic of China}
\affiliation{School of Astronomy and Space Science, University of Science and Technology of China, Hefei 230026, People's Republic of China}

\author{Xianzhong Zheng}
\affiliation{Purple Mountain Observatory, Chinese Academy of Sciences, Nanjing 210023, People's Republic of China}




\begin{abstract}
As the time-domain survey telescope of the highest survey power in the northern hemisphere currently, Wide Field Survey Telescope (WFST) is scheduled to hourly/daily/semi-weekly scan northern sky up to $\sim 23$~mag in four optical ($ugri$) bands.
Unlike the observation cadences in the forthcoming regular survey missions, WFST performed ``staring'' observations toward Galactic plane in a cadence of $\approx 1$~minute for a total on-source time of about 13 hours, during the commissioning and pilot observation phases.
Such an observation cadence is well applied in producing densely sampling light curves and hunting for stars exhibiting fast stellar variabilities.
Here we introduce the primary methodologies in detecting variability, periodicity, and stellar flares among a half million sources from the minute-cadence observations, and present the WFST $g$-/$r$-band light curves generated from periodic variable stars and flaring stars.
Benefit from high photometric precisions and deep detection limits of WFST, the observations have captured several rare variable stars, such as a variable hot white dwarf (WD) and an ellipsoidal WD binary candidate.
By surveying the almost unexplored parameter spaces for variables, WFST will lead to new opportunities in discovering unique variable stars in the northern sky.

\end{abstract}



\section{Introduction} \label{sec:intro}
\begin{table*}
\centering
\footnotesize
\caption{Basic information for WFST minute-cadence observations.
\label{tab:obs_info}
}
\begin{tabular}{lcccccccc}
\hline\hline
Observation & Target & Field direction & $N_{\rm epoch}$  &Filter &   Exposure (s) & Start Time (UTC) & End Time (UTC) \\
\hline
GP-20230918$^*$ & J0526+5934 & $05:26:10.43+59:34:45.1$ &  165 &  $g$ &20 & 2023-09-18 18:20:05 & 2023-09-18 21:20:29 \\
GP-20231116 & J0526+5934 & $05:26:10.43+59:34:45.1$ &  254 &  $g$ & 20 & 2023-11-16 20:03:47 & 2023-11-16 23:22:38 \\
GP-20240206$^+$ & Platais~3 & $04:39:54.24+71:16:48.0$ &  130 &  $g$  & 20 & 2024-02-06 12:06:36 & 2024-02-06 15:54:46 \\
GP-20240209 & Platais~3 & $04:39:54.24+71:16:48.0$ &  258&  $g$ & 20 & 2024-02-09 12:10:52 & 2024-02-09 15:43:02 \\
GP-20240509 & Collinder~350 & $17:48:04.32+01:31:30.0$ &  176&  $r$  & 30 & 2024-05-09 17:57:13 & 2024-05-09 20:45:26 \\
\hline
\end{tabular}
\tablecomments{$^*$Because the operation control system (OCS, \citealt{Zhu+etal+2024+WFST_ocs}) started to work since 2023 October 30th, the observation GP-20230918 was executed manually.\\
$^+$Due to technical tests, GP-20240206 was occasionally interrupted for 73.3 minutes.
}
\end{table*}

Over the last decade, numerous frontier topics on fast stellar variabilities with time scales of minutes to hours have emerged. They involve ultracompact binaries (UCBs, \citealt{Burdge+etal+2019+7min,Burdge+etal+2020+systematic_VBs,lin+etal+2024+sdB_binary_NatAs}), blue large-amplitude pulsators (BLAPs, \citealt{Pietrukowicz+etal+2017+BLAP_NatAs,lin+etal+2023+BLAP_NatAs}), rapidly rotating magnetic white dwarfs (WDs; \citealt{Caiazzo+etal+2021+moon,Williams+etal+2022+MWD}), transitional cataclysmic variables (CVs; \citealt{Burdge+etal+2022+TCV+nature}), black widows \citep{Burdge+etal+2022+Nature+widow}, and fast flaring stars \citep{Gunther+etal+2020+flares,Aizawa+etal+2022+fast_flare,Howard+etal+2022+flare_morphology,Liu+etal+2023+tmts_flaring_stars}. 
The fast stellar variabilities imply extraordinary physical conditions in pulsation, rotation, orbital motion, or magnetic field for these variable stars, and thus provide crucial windows to study the stellar physics under extreme conditions.

UCBs are a class of binaries with orbital periods shorter than $\sim 70$~min. Due to their compact orbits, UCBs avoid to harbor a main-sequence component star \citep{Rappaport+etal+1982+compact_binary}, and is thus composed of neutron star (NS; \citealt{Lin+Yu+2018+UCXBs,Wang+etal+2021+UCXBs}), WD \citep{Burdge+etal+2020+systematic_VBs,Ren+etal+2023+UCBs}, or hot subdwarf \citep{Geier+etal+2013+hsb,Kupfer+UCB+hsb_2020,Kupfer+UCB+hsb_2020_two,Finch+etal+2022+GW_vbs,lin+etal+2024+sdB_binary_NatAs}.
A part of UCBs are predicted to generate strong gravitational wave (GW) radiation in millihertz (mHz) passband and are thus detectable from space-borne GW detectors, e.g. Tianqin \citep{Huang+etal+2020+tianqin} and Laser Interferometer Space Antenna (LISA, \citealt{LISA+2017}).
Since most UCBs are unresolved in the GW detectors and their GW signals are superposed incoherently, millions of UCBs in our Galaxy will produce a stochastic foreground signal (i.e. confusion noise, \citealt{Hils+etal+1990+GW,Ruiter+etal+2010+GW_foreground}).
Fortunately, beyond the confusion noise, space-borne GW detectors will be able to resolve tens of thousands of the UCBs \citep{Nelemans+etal+2001+population,Lamberts+etal+2019+population,Korol+etal+2022+population,Amaro-Seoane+etal+2023+LISA}.
Nowadays, time-domain survey telescopes are capable of searching for these ``verification''  UCBs individually from electromagnetic observations in advance, and thus guarantee the operations of the GW detectors \citep{Kupfer+etal+2018+VBs,Kupfer+etal+2024+LISA,Finch+etal+2022+GW_vbs}.

BLAPs represent a rare class of short-period, large-amplitude hot pulsating stars \citep{Pietrukowicz+etal+2017+BLAP_NatAs}. 
Since the BLAPs were first discovered with the Optical Gravitational Lensing Experiment (OGLE), dozens of BLAPs (or candidates) have been identified successively \citep{Kupfer+2019+high-g_BLAPs,Lin+etal+2021+tmtsI,McWhirter+Lam+2022+blap_candidates,Pigulski+Kolaczek-Szymanski+2022+TESS_BLAP,Borowicz+etal+2023+outer_Bulge_BLAPs,Borowicz+etal+2023+disk_BLAPs,Chang+etal+2024+SMSS_BLAP}.
So far, three candidate stellar models have been proposed for understanding the physical origin of BLAPs: helium-core pre-white dwarfs (pre-WD; \citealt{Corsico+etal+2019+book+pulsatingWD,Byrne+etal+2020+faint_BLAP,Byrne+etal+2021+population_BLAP} ), core helium-burning (CHeB) subdwarfs \citep{Byrne+etal+2018+BLAPs,Wu+etal+2018+CHeB_BLAP,Meng+etal+2020+SN_BLAP}, and shell helium-burning (SHeB) subdwarfs \citep{Xiong+etal+2022+SHeB,lin+etal+2023+BLAP_NatAs}. 
However, due to deficient and in-homogeneous samples,  the physical origin of BLAPs is still controversial.
As introduced by \cite{lin+etal+2023+BLAP_NatAs}, BLAPs may have diverse origins, and the rates of period changes play a crucial role in diagnosing their natures. 
With the operations of large-aperture survey missions,
the growing  BLAP samples and their monitoring data will essentially improve the understanding of these mysterious hot pulsating stars.

Stellar flares are dramatically explosive phenomena triggered by impulsive magnetic reconnection in the corona, potentially accompanied by coronal mass ejections (CMEs, \citealt{Argiroffi+etal+2019+NatAs+flare_CME}), relativistic electron beams \citep{Reid+etal+2014+typeIII_beam}, and chromospheric evaporation \citep{Gudel+eatl+2002+Chrom_evaporation}.
However, since the typical timescales of optical flares are from minutes to hours, the daily-/hour-cadence survey missions fail to reproduce temporal flare structures, leading to incorrect estimations for the amplitudes, durations, and energies of flares \citep{Yang+etal+2018+SC_LC_flares}.
Hence, the minute-cadence observations, especially executed by large-aperture, wide-field instruments, will provide rare opportunities to glimpse the flare profiles from those dark dwarf stars.
In the future, the ultrashort observation cadence ($\approx$ 0.3 second) of Tianyu project \citep{Feng+etal+2024_tianyu} even allows us trace the rapid rise phases of flares and thus helps understand the detail process of magnetic energy release in dwarf stars.

\begin{figure*}
    \includegraphics[width=0.49\textwidth]{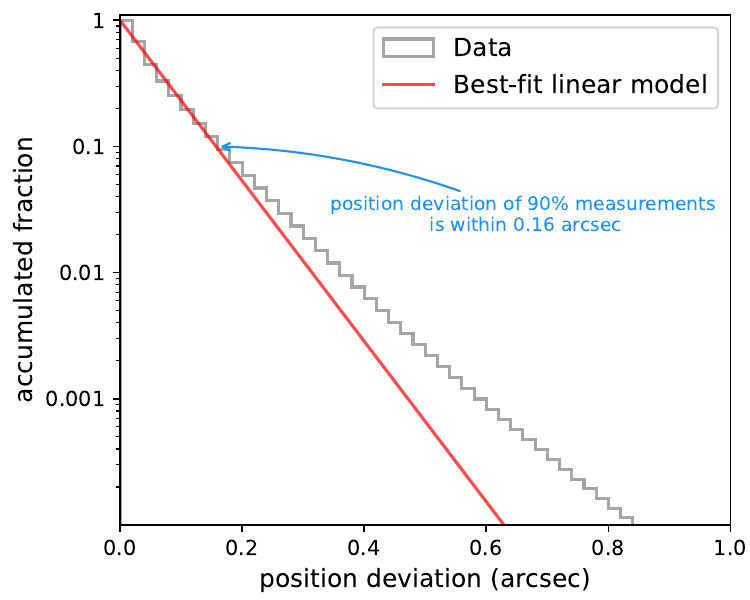}
    \includegraphics[width=0.49\textwidth]{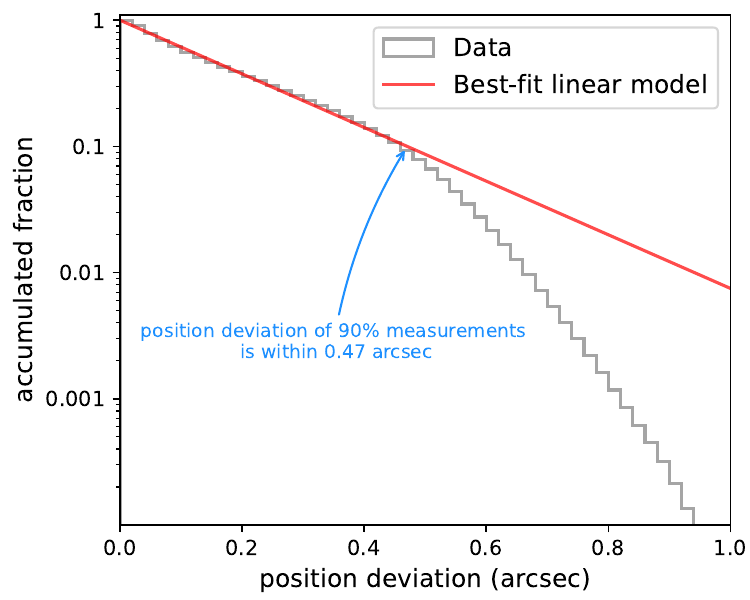}
    \caption{
    Accumulated fractions for the position deviations given from the minute-cadence observations on February 9th, 2024. 
      \emph{Left:} Position deviations anchored at average coordinates of position measurements.
      \emph{Right:} Position deviations anchored at the coordinates from the first detections of sources.
      The bin width is 0.02~arcsec, and the red lines represent best-fit linear models for the bins with an accumulated fraction $> 10\%$.
     } 
    \label{fig:location_deviations}
\end{figure*}

Aiming at hunting the fast stellar variabilities in Galaxy, a few wide-field instruments started performing high-cadence survey missions,  e.g. Zwicky Transient Facility (ZTF, \citealt{ZTF+2019+products,ZTF+2019+first}) high-cadence Galactic plane survey \citep{Kupfer+etal+2021+high_candece},  OmegaWhite \citep{Macfarlane+etal+2015+omega_I,Toma+etal+2016+omegaII}, and Tsinghua University--Ma Huateng Telescopes for Survey (TMTS, \citealt{Zhang+etal+2020+tmts_performance,Lin+etal+2021+tmtsI,Guo+eatl+2024+periodic_variables}).
Benefit from densely sampling photometry, these missions can well capture the fast light variations from diverse variable stars. 
To investigate rapid stellar variabilities in deep fields, we performed a minute-cadence Galactic plane survey using the Wide Field Survey Telescope (WFST, \citealt{Lou+etal+2016+WFST_design,Wang+etal+2023+WFST_science}), a 2.5-meter-aperture telescope at Lenghu Observatory with a median seeing $0.75\arcsec$ \citep{Deng+etal+2021+lenghu}.
Equipped with a 6.5-deg$^2$ field of view (FoV) and 0.765-gigapixel mosaic CCD camera \citep{Zhang+etal+2024+WFST_CCD,Geng+etal+2024+WFST_ccs,Feng+etal+2024+WFST_electronics}, WFST's design enables efficient discovery of transients \citep{Lin+etal+2022+TDE,Liu+etal+2023+kilonova,Huang+etal+2024+TDE}, variables, and solar system objects \citep{Lu+etal+2025+NEO,Wang+etal+2025+HOO} down to $g,r \approx 23$~mag \citep{Lei+etal+2023+WFST_limiting, wan2025pilotsurveyglobularclusters}.
In this paper, we present an overview for the WFST minute-cadence Galactic plane survey executed in its commissioning observation phase (2023.09 -- 2024.02) and pilot survey phase (2024.03 -- 2024.06). 


\section{Observations} \label{sec:obs}


Since the commissioning of WFST on 2023 September, we have performed  minute-cadence observations on Galactic plane for 5 times, with a total duration of about 13~hours.
As shown in Table~\ref{tab:obs_info}, these observations were executed uninterruptedly for 3--4 hours, except for the observation GP-20240206 being occasionally interrupted for 73.3 minutes \footnote{Because the interruption time is far shorter than a day, we still refer to GP-20240206 as an uninterrupted observation throughout this paper.}.
With a dead time (caused by overhead) of about 30 seconds, the observation cadences are $\approx$50s and $\approx$60s for 20s-exposure and 30s-exposure campaigns, respectively.
Notice that, since GP-20230918 was executed manually, the cadence in this observation is non-uniform in an average cadence of about 61~seconds.
All these observations covered 3 non-overlapping fields (i.e. J0526+5934, Platais~3 and Collinder~350), corresponding to a total unrepeated area of $\approx 20$~deg$^2$. Among them, J0526+5934 and Platais~3 were visited twice for capturing general periodic variable stars\footnote{In this work, we refer to the variable stars with period longer than those typical short-period variable stars as general periodic variable stars, e.g. EW eclipsing binaries and RR Lyrae stars.} , while Collinder~350 was simultaneously observed with both WFST and Australian Square Kilometre Array Pathfinder (ASKAP; \citealt{Murphy+etal+2013+VAST,Murphy+etal+2021+VAST,Wang+etal+2023+ASKAP}).
All these observations were executed using $g$- or $r$-band filter.
The effective wavelengths of the $g$- and $r$-band filters are 476.34~nm and 620.57~nm in theory \citep{Lei+etal+2023+WFST_limiting}, respectively, which are similar to those of Sloan Digital Sky Survey (SDSS, $\lambda_{\rm eff, SDSS-g}=467.18$~nm and $\lambda_{\rm SDSS, SDSS-r}=614.11$~nm, \citealt{Fukugita+etal+1996+SDSS,York+etal+2000+SDSS}) but with higher transmittance.

The WFST raw data were primarily dealt with the modified version of the LSST pipeline \citep{2017ASPC..512..279J}, which is developed for the general purpose of the photometric data reduction. For the single frame processing, the WFST pipeline (see \citealt{Cai+etal+2025+pipeline}) contains several basic steps. First, the instrumental signatures were removed with the corresponding calibration exposures, including bias, (sky) flats, fringe and cross-talk. The instrumental-signatures-removed images were then labeled pixel-by-pixel with blending, saturation, cosmic ray contamination etc, indicating which part of the images should be ignored or carefully considered in the further analysis. 
The background was estimated as the median value of each $512 \times 512$ pixel sub-region on the image. A 2-dimensional $6^{\mathrm{th}}$-order Chebyshev polynomial was applied in both the $X$ and the $Y$ directions to derive a smooth distribution of the background, and finally produced a background image and a background-subtracted science image.
Sources above the detection thresholds were then detected within the background-subtracted science images, 
and images' point-spread-functions (PSF) were estimated with a selected sub-sample of the sources. The flux of the sources were measured in both diverse apertures (4.5--70~pixels) and in PSF modeling, which were finally calibrated with external references---the astrometric parameters are calibrated to the Gaia DR3 \citep{Gaia_Collaboration+2016+performance,Gaia+DR3+2022} and the $g$-/$r$-band photometric parameters are calibrated to PanStarr DR2 \citep{Chambers+etal+2016+PanSTARRS1,Magnier+etal+2020+PS1_Calibration,Flewelling+etal+2020+PS1_products}. 
Throughout this work, all photometric fluxes were calibrated to AB magnitudes, and the times corresponding to exposure midpoints were converted into barycentric modified Julian dates with barycentric dynamical time ($\rm BMJD_{TDB}$).



All WFST photometric measurements labeled with blending, saturation, cosmic ray, and closing to the image edge were discarded throughout this work.
By cross-matching the photometric measurements within a radius of 1~arcsec, we extracted light curves for all sources within the fields. 
For checking whether the 1-arcsec radius is large enough to bind all photometric measurements from same sources, we additionally investigate astrometric deviations for all WFST sources with epochs more than 20.
The deviations were given by the angular separations between each position measurements and anchored coordinates, i.e. the average coordinates of sources here.
As shown in the left panel of Fig.~\ref{fig:location_deviations}, the accumulated fractions suggest that, the astrometric precision of WFST is better than 0.16~arcsec for 90\% measurements in these early observations.
However, since we tend to take the position measurements from first detections as the coordinates of sources, rather than the average ones, a larger radius should be adopted in cross-matching (see the right panel of Fig.~\ref{fig:location_deviations}).
Here the 1-arcsec radius ensures an $\approx$0.01\% possibility at most that a measurement from a source cannot match its first detection, which effectively prevents splitting a source into multiple sources.


\section{Methods} \label{sec:meth}

\begin{figure*}
    \includegraphics[width=0.49\textwidth]{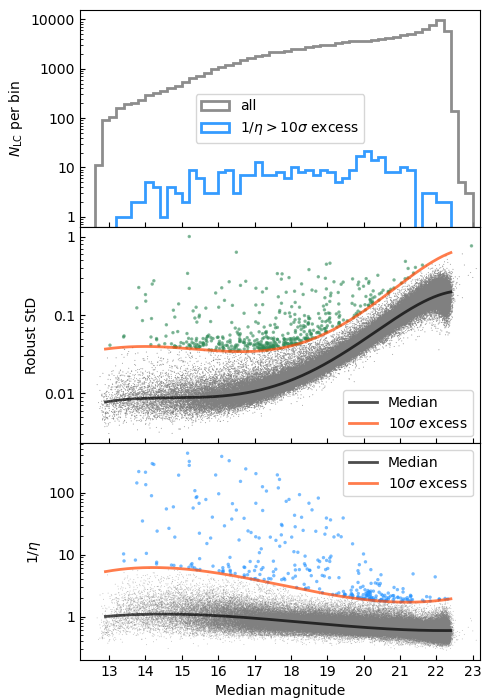}
    \includegraphics[width=0.49\textwidth]{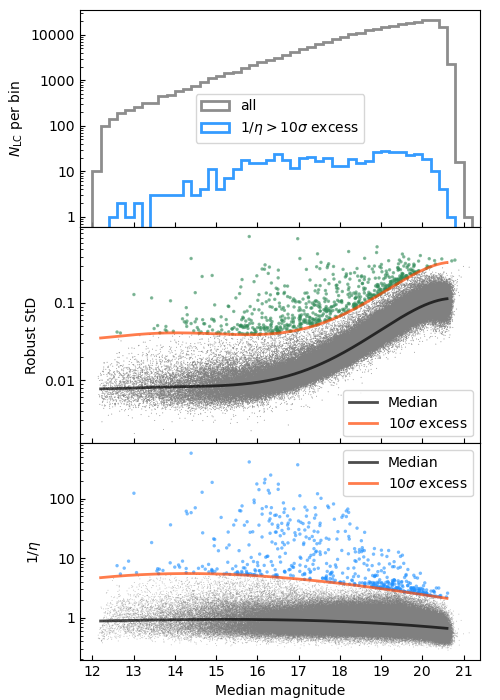}
    \caption{Distributions for number (upper), robust standard deviation (middle) and inverse von Neumann ratio (lower) against the median AB magnitude, derived from the light curves extracted from $g$-band observations GP-20240209 (left) and $r$-band observations GP-20240509 (right).
      \emph{Upper panels:} The grey lines represent the number of light curves (with $\geq 20$ epochs) per bin, while the blue lines  represent the number of candidate variable stars (with $1/\eta$ above the $10\sigma$ threshold) per bin. The bin size is 0.2~mag.
      \emph{Middle panels} \& \emph{Lower panels}:
      The black and red lines represent the medians and $10\sigma$ excesses for the distributions of variability indices, respectively.
      The grey and green/blue points represent the light curves with variability indices below and above the $10\sigma$ thresholds, respectively.
     } 
    \label{fig:variability_detection}
\end{figure*}

\begin{table*}
\centering
\footnotesize
\caption{Statistical information for WFST minute-cadence observations.
\label{tab:statis_info}
}
\begin{tabular}{lccccccccccccc}
\hline\hline
Observation & Average FWHM$^a$ & \multicolumn{3}{c}{Photometric precision$^b$ (mag)} & \multicolumn{2}{c}{Limiting magnitude$^c$ (mag)}   & \multicolumn{2}{c}{Number of sources} \\
\cline{3-5}
 & (arcsec) & at 15~mag & 18~mag & 20~mag  &  5$\sigma$&  10$\sigma$ &  20 epochs &  50 epochs\\
\hline
GP-20230918 & 2.00 & 0.0057& 0.0138& 0.0547 & -$^d$ & 20.859& 76,517 & 60,635\\
GP-20231116 & 1.81 & 0.0124& 0.0181 & 0.0604 & 21.786& 20.716 & 115,637 & 95,847\\
GP-20240206 & 2.51 & 0.0082& 0.0196 & 0.0825 & 21.371& 20.300 & 76,773 & 57,436\\
GP-20240209 & 1.67 & 0.0088& 0.0147 & 0.0480 & 22.338& 21.071 & 119,199 & 98,475 \\
GP-20240509 & 2.40 & 0.0084& 0.0214 & 0.0909 & -$^e$ & 20.267 & 262,628 & 219,339 \\
\hline
\end{tabular}
\tablecomments{
$^a$ The average full widths at half maxima (FWHMs) are determined by point spread functions of field sources, which characterize the observation conditions.\\
$^b$ Photometric precision values are calculated by averaging robust standard deviations (StDs) binned at 15, 18, and 20~magnitudes.  \\
$^c$ Limiting magnitudes are defined as the median magnitudes at robust StD$=0.2$~mag (5$\sigma$ level) and $=0.1$~mag (10$\sigma$ level). \textbf{Notice that, the minute-cadence observations are confined to a single night and a localized sky region; consequently, the derived limiting magnitudes here are not representative of WFST's full capabilities.} \\
$^{d, e}$ The implementation of a stringent 5$\sigma$ detection threshold (see Section~\ref{Sec:variability}) inherently precludes determination of their 5$\sigma$ limiting magnitudes through quantification of photometric scatter, as faint sources below the threshold are systematically excluded from the analysis.
}
\end{table*}

In order to select diverse variables from millions of WFST light curves, we applied three different algorithms in the light-curve analysis (LCA, \citealt{Sokolovsky+etal+2017+variables} ), namely von Neumann ratio $\eta$, Osten's method \citep{Osten+etal+2012+flare_archive}, and Lomb–Scargle periodogram (LSP, \citealt{Lomb+1976,Scargle+1982}). Since \cite{Lin+etal+2021+tmtsI} has detailed how these algorithms worked on the uninterruptedly light curves obtained from minute-cadence survey observations, here we introduce only the basic principles for these algorithms and their performances for WFST observation data.
Furthermore, we also cross-match the WFST sources with Gaia~DR3 catalog \citep{Gaia_Collaboration+2016+performance,Gaia+DR3+2022} for locating them in color-magnitude diagram (CMD), which plays a crucial role in preliminary identifications of variable stars.

\subsection{Variability detection}
\label{Sec:variability}

In order to test the variability detection in the observation data, we extracted light curves (with $\geq$20~epochs) from $g$-band observation GP-20240209 and $r$-band observation GP-20240509.
Among them, GP-20240209 contributed 119,119 uninterrupted light curves, while GP-20240509 provided 262,628 ones.
The statistical characteristics for these light curves are presented in Fig.~\ref{fig:variability_detection}.
In the $g$-band observations, the WFST light curves are distributed dominantly from 12.9 to 22.3 mag, and the number of light curves increases along with the median magnitude and peaks around 22~mag. 
Under suboptimal observation conditions (see Table~\ref{tab:statis_info}), the median magnitudes of $r$-band light curves here range from 12.2 to 20.6 mag.
Notice that, varying source detection thresholds were implemented in different observation phases: $3\sigma$ above background noise for GP-20231116, GP-20240206, and GP-20240209, versus 5$\sigma$ for GP-20230918, GP-20240509, and subsequent observations. While the stringent threshold excludes marginally detectable faint sources, it effectively suppresses artificial detections from observations.

Similar to the standard deviation (StD), robust StD is a variability index quantifying the scatter of brightness measurements based on 25 percentile of the measurements above/below the median.
Because only inner 50 percentile of the measurements are used to derive the StD (see also \citealt{Ofek+ZTF+DR1+2020}), robust StD is immune from outliers or occasional real variations (e.g. fast flares).
As shown in the middle panels of Fig.~\ref{fig:variability_detection}, the robust StD raises from 0.008 mag to $\sim$0.2~mag with the darkening of sources, while the variation trend represents the photometric precisions at various magnitudes.
Based on the robust StDs binned at various magnitudes, the statistical information for WFST minute-cadence observations are summarized in Table~\ref{tab:statis_info}.

The light curves that present significantly larger scatters than the expected uncertainties, are likely generated from variable sources.
By fitting the 10$\sigma$ excesses at various magnitudes with $5^{\mathrm{th}}$-order polynomial models, 462 candidate variable sources and 634 ones were selected from GP-20240209 and from GP-20240509, respectively, implying about 3 candidate variables among per 1,000 observed sources.
However, due to blending, saturation, stray lights, and varying atmospheric extinction with the altitude, the candidate variable sources inevitably include a large proportion of non-astrophysically variable sources\footnote{The sources with light variations that are not caused by real astrophysical variability.}.
As \cite{Kupfer+etal+2021+high_candece} introduces, about 60\%-85\% of the candidate variable sources present spurious variations from the ZTF high-cadence observations, especially in those crowded fields; \cite{Lin+etal+2021+tmtsI} also revealed a false positive rate\footnote{The ratio of the non-astrophysically variable sources to the all candidate variable sources here.} of 67\% from their 10$\sigma$-excess variable candidates in the first-year minute-cadence observations of TMTS.
By visually checking the WFST light curves from all candidate variable sources, we found that the false positive rate is 76\% for GP-20240209 and 65\% for GP-20240509, respectively.
However, because robust StD is extremely insensitive to the temporary light variations in the light curves, we tended to adopt another variability index to detect the brightness variability of sources including those flaring stars.

By comparing 18 variability indices that quantify scatter and/or correlation among photometric time series, \cite{Sokolovsky+etal+2017+variables} concluded that von Neumann ratio $\eta$ (or inverse von Neumann ratio $1/\eta$, conventionally) is one of the best indices in selecting candidate variables.
The von Neumann ratio can test the independence of successive brightness measurements, and is thus applied in variability detection for light curves affected by outliers.
The $1/\eta$ can be calculated by 
\begin{equation}
\frac{1}{\eta}=\frac{\sum\limits_{i=1}^{N} (m_{i}-\overline{m} )^2}
{\sum\limits_{i=1}^{N-1}  (m_{i+1}-m_{i})^2 },
\label{Eq:neumann_ratio}
\end{equation}
where $m_{i}$ and $m_{i+1}$ represent the $i$th and ($i+1$)th magnitudes in a light curve, respectively.
$\overline{m}$ is the average magnitude over all epochs, and $N$ denotes the number of epochs.
The $1/\eta$ is expected to be 0.5 for an ideal time series following a Gaussian distribution. However, due to astrophysical variations and various photometric issues, the distributions of real photometric measurements are actually deviated from Gaussian distributions.
As shown in the lower panels of Fig.~\ref{fig:variability_detection},  the $1/\eta$ decreases from $\sim$1 to 0.6 along with the magnitude. 
Similarly, we fitted the 10$\sigma$ excesses against median magnitudes using $3^{\mathrm{th}}$-order polynomial models, which lead to 306 and 527 candidate variable sources selected from GP-20240209 and from GP-20240509, respectively. These candidate variable sources include a dozen of flaring stars that were missed in the selections based on robust StD.
By visually checking the light curves, we found that the false positive rate drops to 60\% for GP-20240209 and 53\% for GP-20240509, respectively, implying that von Neumann ratio is a much better variability index to select variable stars.
As both $r$- and $g$-band minute-cadence observations suggest that about 0.1\% of observed sources exhibit real astrophysical variability, WFST is expected to reveal several million variable sources among five billion significantly detected sources within Galaxy in the future \citep{Wang+etal+2023+WFST_science}.

Furthermore, in order to search for specific types of variable stars, we prefer to use Osten’s method and LSP to select flaring stars and  periodic variable stars, respectively, while the variability indices are used as an auxiliary tool to improve the completeness of selections.

\subsection{Flare detection}

\begin{figure}
    \includegraphics[width=0.47\textwidth]{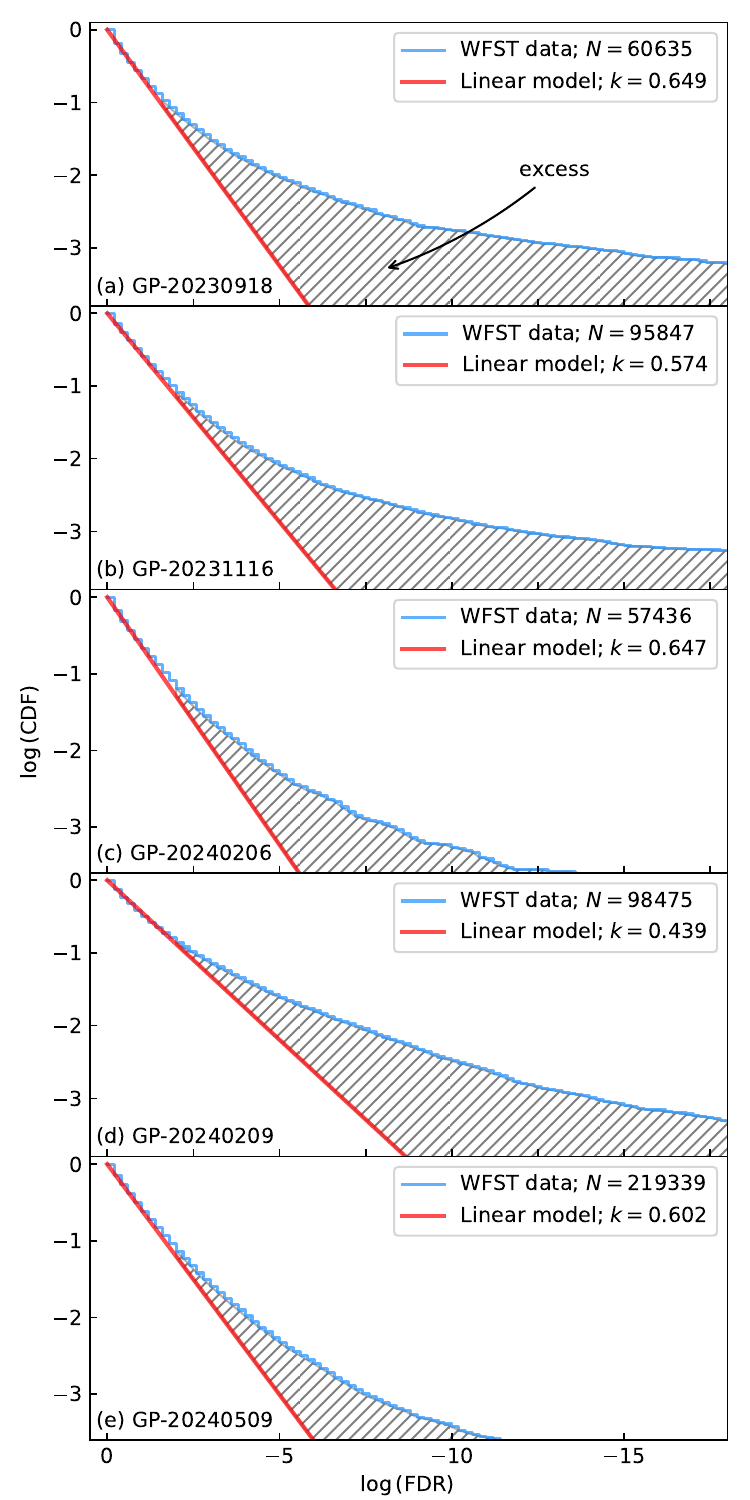}
    \caption{Cumulative distribution functions (CDFs) of false discovery rates (FDRs) for flare detection from WFST minute-cadence observations.
    The FDR histograms (blue solid lines) in panel~a--e are given by $\phi_{\rm VV}$ sequences derived from five WFST minute-cadence observations, respectively.
    And the red solid lines are the best-fit linear models for the histogram bins at $\log ({\rm CDF})\geq -1$.
    The bin size is 0.2 .
    The shaded area indicate that the observed FDR distributions excess the models.
     } 
    \label{fig:flare_fdr_distribution}
\end{figure}

\begin{figure*}
    \includegraphics[width=0.96\textwidth]{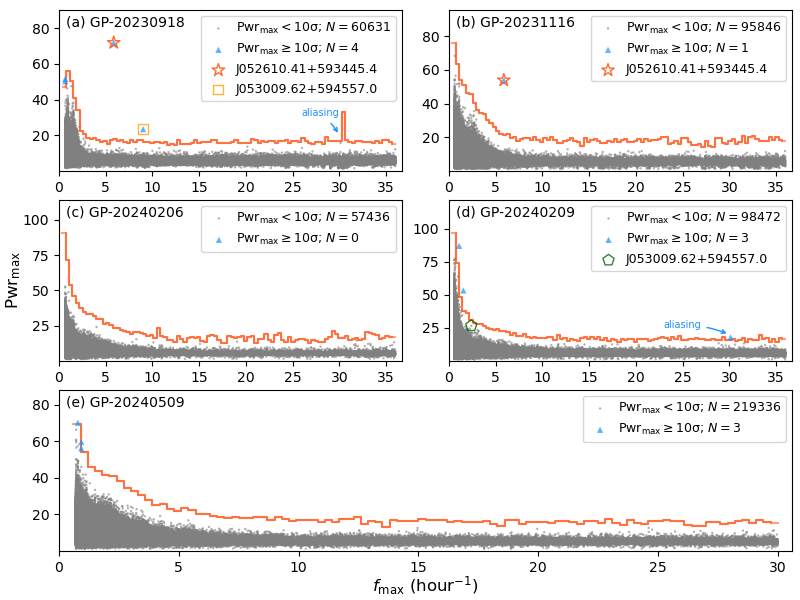}
    \caption{Distributions of maximum power $\rm Pwr_{ max}$ versus their corresponding frequency $f_{\rm max}$. Data points in panel~a--e are given by LSPs derived from five WFST minute-cadence observations, respectively. The red solid line represents 10-$\sigma$ excesses, and the total numbers ($N$) of sources above/below the $10\sigma$ threshold are indicated in the legends.
     Several interesting short-period objects are highlighted with hollow symbols.
     } 
    \label{fig:periodicity_detection}
\end{figure*}

In order to search for fast flares in minute-/hour-timescale from minute-cadence observation data, we applied the flare selection method described in \cite{Osten+etal+2012+flare_archive} and the false discovery rate (FDR) analysis used by \cite{Lin+etal+2021+tmtsI}.
Before the selections, all 531,732 uninterrupted light curves (with epochs $\geq$ 50) were detrended using a compound model of $4^{\mathrm{th}}$-order Fourier series and $2^{\mathrm{nd}}$-order polynomial (see the Eq.~7 in \citealt{Lin+etal+2021+tmtsI}). 
Given the hour-level observations spans, the polynomial components here are intended for offsetting potential long-timescale variations induced by both astrophysical (e.g. long-period eclipsing binaries) and non-astrophysical (e.g. varied altitude angles) factors.

For obeying the normal distribution, the detrended magnitudes ($m_{\rm dtr}= m-m_{\rm model}$ ) are normalized as 
\begin{equation}
V_i= \frac{\overline{m_{\rm dtr}} - m_{{\rm dtr},i}}{\sigma_{\rm dtr}} ,
\label{Eq:normalized_residuals}
\end{equation}
where $m_{{\rm dtr}, i}$  represents $i$th detrended magnitude. 
$\overline{m_{\rm dtr}}$ and $\sigma_{\rm dtr}$ represent the median and robust standard deviation of detrended magnitudes, respectively. 

Following the approach introduced by \cite{Osten+etal+2012+flare_archive}, the flares in light curves could be detected through at least two consecutive positive outliers in the time series.
The index quantifying the significance of two consecutive outliers in a time series is defined as the product of continuous two normalized detrended magnitudes, namely
\begin{equation}
 \phi_{{\rm VV}, i}=V_{i} \times V_{i+1}~.
 \label{Eq:phiVV}
\end{equation}
For the purposes of selecting flares (rather than dips or eclipses), the adopted $\phi_{{\rm VV}, i}$ pairs require both $V_{i}>0$ and $V_{i+1} > 0 $.
To simplify calculations, only the maximum $\phi_{{\rm VV}, i}$ over a time series (i.e. $\phi_{\rm VV,max} = \max \limits_i \{ {\phi_{{\rm VV}, i}\,|\,_{V_{i}>0, V_{i+1}>0} } \}$ ) is applied to the flare detection.
In other words, a light curve having a higher $\phi_{\rm VV,max}$  is more likely to present flares.

Benefit from the probability density function (PDF) for $\phi_{\rm VV}$ given by \cite{Lin+etal+2021+tmtsI}, we can calculate the FDRs of flares in theory for WFST light curves via
\begin{equation}
{\rm FDR}=1-[1 - \frac{1}{2\pi} \int^{\infty}_{ \phi_{\rm VV, max}  } \,K_0(\vert x \vert) \,{\rm d}\,x    ]^{N-1}~,
\label{Eq:FDR}
\end{equation}
where $K_0$ represents the special ($n=0$) case of \emph{modified Bessel function of the second kind} \citep{Abramowitz+Stegun+1972+mathematical+function}, and $N-1$ is the number of $\phi_{\rm VV}$ for a time series of $N$ epochs.
The cumulative distribution functions (CDFs) of these FDRs are shown in Fig.~\ref{fig:flare_fdr_distribution}.
Given the prerequisite that only a small fraction of light curves exhibit flares, we fitted the histogram bins at $\log ({\rm CDF})\geq -1$ (i.e. the cumulative distributions for 90\% light curves) with a linear model, $\log ({\rm CDF})=k\times \log ({\rm FDR})$.
Since the magnitudes in light curves do not tightly follow Gaussian distributions and are not completely independent of each other, it is common for the cumulative distributions of FDRs to deviate from the ideal null distribution (i.e. ${\rm CDF}={\rm FDR}$, see also \citealt{Lin+etal+2021+tmtsI}).
By regarding the best-fit models as the null distributions, FDRs can be calibrated via the best-fit slope $k$, namely ${\rm FDR}_{\rm cal} = {\rm FDR}^k$ .
Given the non-astrophysical outliers induced by potential issues (e.g. bleeding of saturated sources) in early observation data, we adopted  a tight threshold for false detection rates (i.e. ${\rm FDR}_{\rm cal} \leq 10^{-6} $), leading to 486 flaring star candidates selected from the light curves. 
By visually checking all candidates, we found that, 33 ones present real flares, while the most of artificial flares are arisen by the bleeding of saturated sources or the incorrect models for detrending.
Nonetheless, the number of flaring stars revealed from the total 15-hour WFST observations is comparable to all flaring stars discovered among first-year TMTS observations \citep{Lin+etal+2021+tmtsI, Liu+etal+2023+tmts_flaring_stars}. 

However, since the observation spans are only 3--4 hours, the Osten's method failed to model the quiescence fluxes for the ``long-duration'' flares lasting for $\gtrsim 1$~hour, leading to the miss of these ``energetic'' flares from the selections above.
Given that the ``energetic'' flares must exhibit significantly strong variability, 
we visually checked the variable star candidates with high $1/\eta$ values and picked out four additional flares.
These ``energetic'' flares having diverse light-curve morphologies are presented in following section.


\newpage

\subsection{Periodicity detection}

Lomb–Scargle periodogram (LSP, \citealt{Lomb+1976,Scargle+1982}) is a common tool to reveal periodic variation signals from unevenly sampling time series.
In order to reveal both short-period variable stars and general periodic variable stars, the LSPs for WFST light curves were calculated under two different sets.

\subsubsection{Short-period variable stars}
\label{sec:method_short_period}
Aiming at searching for short-period variable stars, we calculated variance-scaled LSPs (see Eq.~1 of \citealt{Coughlin+etal+2020+ZTFprojectionII} or Eq.~5 of \citealt{Lin+etal+2021+tmtsI}) for all uninterrupted light curves with epochs $\geq 50$ within frequency from $2/T$ to $f_{\rm nyq}$, where $T$ represents the time spans of uninterrupted observations and $f_{\rm nyq}$ is the (pseudo-)Nyquist frequency  \citep{VanderPlas+2018+LSP}.
The $f_{\rm nyq}$ here was set to a half of average sampling rate $f_0$, i.e. 1/100~Hz for 20s exposures and 1/120~Hz for 30s exposures, respectively.
Due to the limitation from the lowest frequency $2/T$, these LSPs can only be used to detect variable stars with a period shorter than 1.5-2 hours.

The false alarm probability (FAP) that quantifies the significance of a periodic signal is generally inferred from the maximum power ($\rm Pwr_{ max}$) over a LSP (e.g. Eq.~54 of \citealt{VanderPlas+2018+LSP}). However, for the uninterrupted light curves obtained within single nights, the distributions of $\rm Pwr_{ max}$ are severely affected by window functions and are thus frequency-dependent. 
To avoid missing short-period variable sources, \cite{lin+etal+2023+tmtsII} suggested that ${\rm Pwr}_{\rm max}$-$f_{\rm max}$ diagram should be applied in the periodicity detection for single-night observations, where $f_{\rm max}$ is the frequency corresponding to the maximum power $\rm Pwr_{ max}$ within a LSP.
As shown in Fig.~\ref{fig:periodicity_detection}, $f_{\rm max}$ distributes over the investigated frequency ranges, while $\rm Pwr_{ max}$ integrally tend to much higher values at the low-frequency ends.
As the targeted star in both GP-20230918 and GP-20231116, the periodicity of J0526+5934 (highlighted in the panel $a,b$ of Fig.~\ref{fig:periodicity_detection}, see also \citealt{Kosakowski+etal+2023+J0526_EW,lin+etal+2024+sdB_binary_NatAs,Rebassa-Mansergas+etal+2024+J0526_WD}) is well reproduced in WFST observations.
Thanks to lower ``noise'' at higher frequency in the ${\rm Pwr}_{\rm max}$-$f_{\rm max}$ diagram, a 6.7-min variable star J053009.62+594557.0 (right ascension $\alpha=82.5401$ and declination $\delta=59.7661$, hereafter J0530+5945,  highlighted in the panel $a$ of Fig.~\ref{fig:periodicity_detection}) emerged out of sixty thousand candidate sources.

In order to pick out periodic variable source candidates with frequency-dependent thresholds, each ${\rm Pwr}_{\rm max}$-$f_{\rm max}$ diagram was divided into 100 uniform frequency intervals, and each frequency bin independently derives the thresholds corresponding to 10$\sigma$ excesses. 
Given the thresholds, 10 (unrepeated) periodic variable source candidates were selected.
By visually checking all these candidates, 9 (out of 10) ones exhibit real periodicity, while only a candidate is induced by the aliasing periodic signals around two minutes.
We have tried to lower the selection threshold in order to include more candidates. However, the weak periodicity from these additional candidates is very difficult to be identified and confirmed.
Nonetheless, we highlighted an interesting periodic variable star (i.e. J0530+5945) below the 10$\sigma$ excesses from the observation GP-20240209 (see the panel $d$ of Fig.~\ref{fig:periodicity_detection}). 
We will present further classifications for these short-period objects through the CMD in the following chapter.

\subsubsection{General periodic variable stars}
\label{sec:method_general_periodic}

\begin{figure}
    \includegraphics[width=0.47\textwidth]{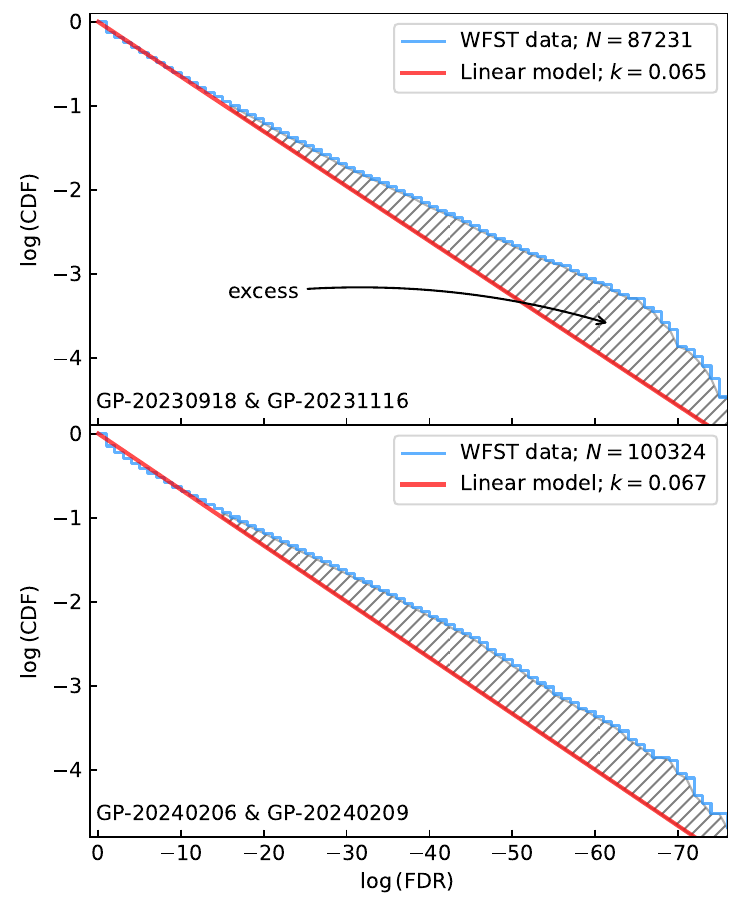}
    \caption{Cumulative distribution functions (CDFs) of false alarm probabilities (FAPs) for periodicity detection from the splicing observations.
    The data sets were obtained from GP-20230918$+$GP-20231116 (upper) and GP-20240206$+$GP-20240209 (lower), respectively.
    The blue lines represent the CDFs derived from WFST observation data, and the red lines are the best-fit linear models for the bins with $\log ({\rm CDF})\geq -1$.
    The bin size is 1.0 .
     } 
    \label{fig:periodicity_fap_distribution}
\end{figure}

In order to mine and present diverse periodic variable stars from our minute-cadence observations,  we spliced the uninterrupted observations having consistent pointings, i.e. GP-20230918+GP-20231116 (SO1 hereafter) and GP-20240206+GP-20240209 (SO2 hereafter).
Since the observation spans are greatly extended to about 3~months/3~days for SO1/SO2, the splicing light curves are thought to be used for detecting periodic variable stars longer than tens of hours.
However, due to the observation strategy developed for capturing short-period variable stars, the splicing observations actually cover minor, non-uniform phase ranges of those potential long-period variable stars.
Owing to extremely uneven sampling cadence, their window functions are dramatically amplified below the frequency of about $3\,{\rm d}^{-1}$.
Hence, we calculated the variance-scaled LSPs for all splicing light curves with epochs $\geq 50$ within a fixed frequency range, $3\,{\rm d}^{-1}$ to $12\,{\rm d}^{-1}$.

The FAPs were obtained following the independent frequency method \citep{VanderPlas+2018+LSP}, namely
\begin{equation}
{\rm FAP}=1-[1-\exp{({\rm -Pwr_{ max} })}]^{N_{\rm eff}}~,
\label{Eq+FAP}
\end{equation}
where $N_{\rm eff}$ is the effective number of independent frequencies, which is assumed to be investigated frequency range in unit of expected frequency width ($\delta f=1/T$). 
As Fig.~\ref{fig:periodicity_fap_distribution} shows, due to the systematic overestimation of  $\rm Pwr_{ max}$ induced by window functions, the cumulative distributions of FAPs are far deviated from the ideal null distribution. 
In order to reproduce the null distributions for the observation data, we tried to fit the histogram bins at $\log ({\rm CDF})\geq -1$ with $\log ({\rm CDF})=k\times \log ({\rm FAP})$. Similar to the approach for flare search, the FAPs were calibrated by ${\rm FAP}_{\rm cal} = {\rm FAP}^k$ based on the best-fit linear models.
By adopting an FAP threshold of 0.1\%, 288 and 374 candidates were selected from SO1 and SO2, respectively.
Among them, we visually confirmed 146 variable stars having real periodic variations.
Notice that, the three-hour uninterrupted observations are not intended to serve the search of the variables in such light-variation periods.
Here mining the periodic variable stars from splicing observations is in an effort to present plentiful types of variable stars in the early results of WFST.

\subsection{Cross-match with Gaia DR3 catalog}

\begin{figure}
    \includegraphics[width=0.47\textwidth]{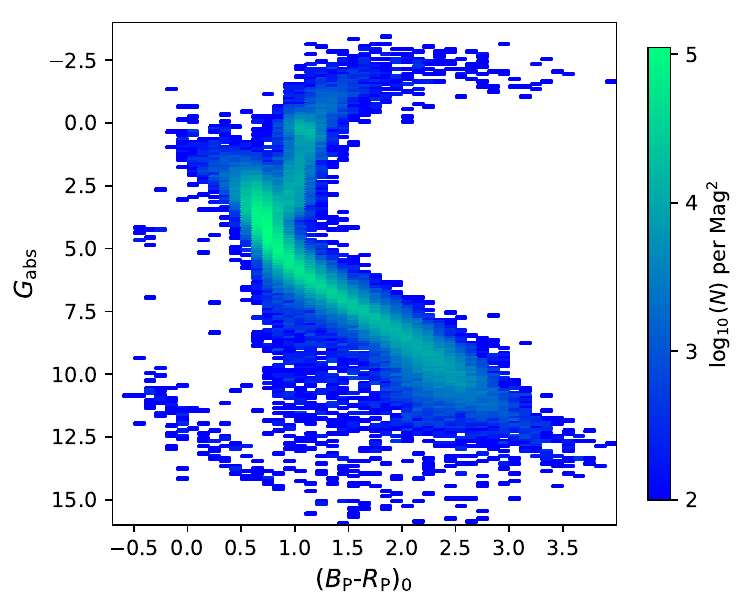}
    \caption{Density distribution of the Gaia DR3-WFST sources across the color-magnitude diagram. 
    Only the Gaia DR3 sources having a reliable parallax measurement are included in the distribution.
    Both magnitudes and colors are calibrated for interstellar dust extinction and reddening.
    The bin size is 0.1$\times$0.1 mag$^2$.
     } 
    \label{fig:cmd_all_sources}
\end{figure}

\begin{figure*}
    \includegraphics[width=0.96\textwidth]{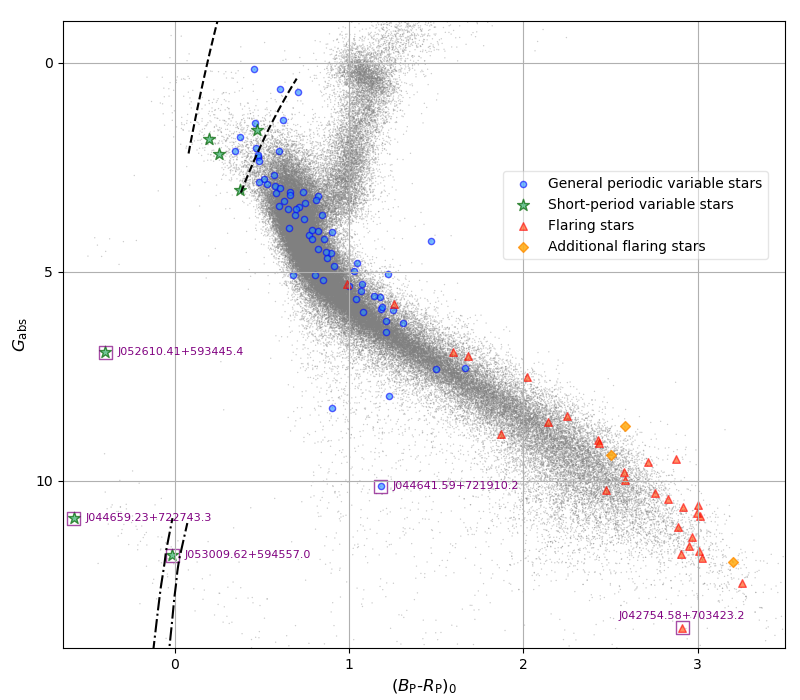}
    \caption{Distributions of periodic variable stars and flaring stars across the CMD.
    The general periodic variable stars, short-period variable stars and flaring stars are selected by three different methods introduced in Section~\ref{sec:meth}.
    The additional flaring stars were discovered by visually checking the variable star candidates with high $1/\eta$ values rather than by the Osten's method.
    Several interesting objects are highlighted by purple squares.
    The grey dots in background are the sources detected from the 15-hour WFST observations. 
    We denote the instability strip edges of $\delta$ Scuti \citep{Murphy+etal+2019+ds} and ZZ Ceti \citep{Caiazzo+etal+2021+moon} with the dashed lines and dot-dashed lines, respectively.
     } 
    \label{fig:periodic_flaring_CMD}
\end{figure*}

Besides the selections based on the diverse light variations, the additional photometric and astrometric information provided from external catalogs is effective for filtering and identifying variable stars.
Since CMD plays a crucial role in the classifications of short-period variable stars \citep{lin+etal+2023+tmtsII}, we cross-matched all 500,460 WFST sources (corresponding to 650,696 uninterrupted light curves having epochs $\geq$ 20 ) with Gaia DR3 sources \citep{Gaia+DR3+2022} within 2~arcsec.
As a result, 489,801 WFST sources (97.87\%) match at least a Gaia counterpart. Among them, 102,509 (20.93\%) Gaia sources have a reliable parallax measurement (i.e. $\varpi/\sigma_\varpi \geq 5.0$ ), which allow us to calculate absolute magnitudes and estimate interstellar dust extinctions.

\begin{figure*}
    \includegraphics[width=0.96\textwidth]{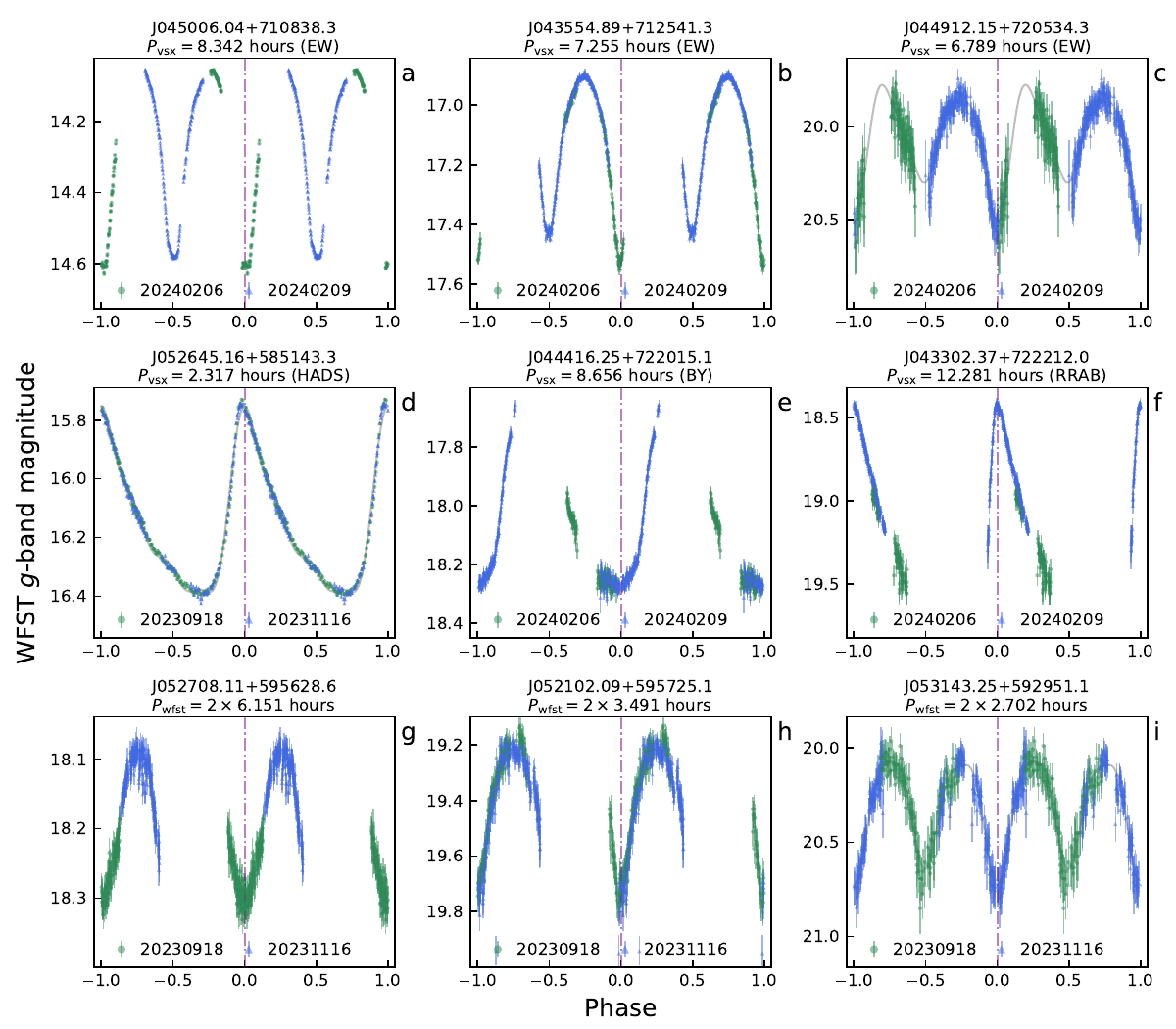}
    \caption{
    A gallery of phase-folded light curves for nine general periodic variable stars revealed by the splicing observations.
    All light curves are obtained from the $g$-band observations of WFST, and the exposure is 20~seconds.
    For the objects identified by VSX, their light curves (panel~a--f) are folded by the period provided from VSX; for the variables first identified by WFST, the light curves (panel~g--i) are folded using twice the photometric periods (i.e. $1/f_{\rm max}$) obtained from WFST LSPs.
    We label the classification type for the objects identified by VSX: EW $=$ EW-type eclipsing binary, HADS $=$ high-amplitude $\delta$ Scuti star, RRAB $=$ RRab Lyrae, and BY $=$ BY Draconis-type variable.
    The light curves having relatively complete phase coverage (i.e. panel c, d and i) are modeled using $4^{\mathrm{th}}$-order Fourier series (the grey solid lines).
     } 
    \label{fig:general_periodic_varibles}
\end{figure*}

We present the density distribution for one hundred thousand Gaia DR3-WFST sources in the CMD (Fig.~\ref{fig:cmd_all_sources}).
Both magnitudes and colors are already calibrated for interstellar dust extinction and reddening, which are obtained from three-dimensional dust map \citep{Green+etal+2019+3dmap} with the \emph{DUSTMAPS Python} package\footnote{\url{ https://github.com/gregreen/dustmaps}} \citep{Green+2018+python}.
As Fig.~\ref{fig:cmd_all_sources} shows, the Gaia DR3-WFST sources cover diverse regions over the CMD, including main sequence, red giant branch, extreme horizontal branch and WD cooling sequence.
Aiming at searching for UCBs harboring double WDs, all WFST sources were also cross-matched with the catalog of WDs in Gaia EDR3 \citep{Gentile_Fusillo+etal+2021+WD_catalog}, leading to 263 WD candidates identified out of the WFST sources. Benefit from the large aperture and thus high detection depth, the number of WDs covered by three pointings of WFST is comparable with the amount covered by 90 TMTS fields \citep{Lin+etal+2021+tmtsI}.

\section{Early results} \label{sec:results}

Following the methods introduced in previous section, we present selected periodic variable stars and flaring stars across the CMD (Fig.~\ref{fig:periodic_flaring_CMD}).
The periodic variable stars occupy the zones of A,F,G-type main-sequence stars while the flaring stars are concentrated on the late-type main sequence.
In this section, we will show WFST performance in detecting the variable stars, and introduced several interesting objects in details.

\subsection{General periodic variable stars}

\begin{figure}
    \includegraphics[width=0.47\textwidth]{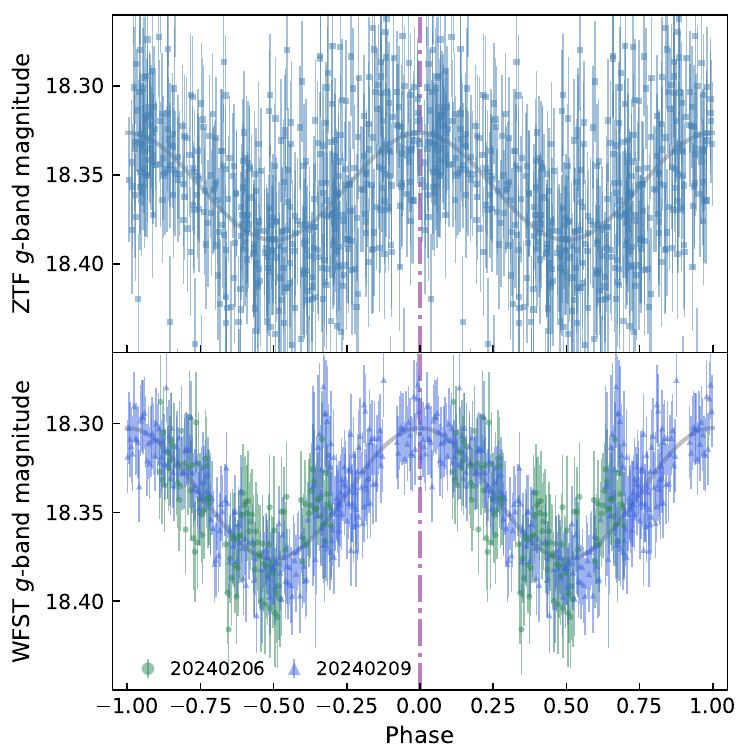}
    \caption{
     Phase-folded light curves of J0446+7219 provided from ZTF DR22 (upper panel) and WFST minute-cadence observations (lower panel).
    Both light curves are folded by the photometric period ($P=2.95126$~hours) derived from the 6-year ZTF $g$-band observations.
    For an ellipsoidal binary, its orbital period is twice the photometric period.
    The grey solid lines represent the best-fit sinusoidal models.
     } 
    \label{fig:periodic_variable_J0446}
\end{figure}

Among 146 selected general periodic variable stars in Section~\ref{sec:method_general_periodic}, 118 objects were already identified by the International Variable Star Index (VSX, \citealt{Watson+etal+2006+VSX}).
According to the variable star types provided from VSX, 89\% of them are identified as EW-type eclipsing binaries and 6\% are RR Lyrae variables.
Notice that, since the typical pulsation period of fundamental-mode (RRab) Lyrae is 0.3--1.0 day, which exceeds the investigated frequency range (i.e. $3\,{\rm d}^{-1}$ to $12\,{\rm d}^{-1}$) of our LSPs. These pulsators were occasionally revealed when the LSPs captured their high-frequency harmonics.
Fig.~\ref{fig:general_periodic_varibles} presents phase-folded light curves for several general periodic variable stars within diverse magnitude ranges.
The photometry with high SNR allows WFST independently discover eclipsing binaries and pulsating stars from 14~mag to darker than 20.5~mag.
Owing to our observation strategy aiming at searching for short-period variable stars, the phases of these ``long-period'' objects are incompletely covered by our observations.
With the coming of regular surveys, we will systematically introduce WFST's detection capability in periodic variable stars in the future.

%

\textbf{A newly discovered ellipsoidal binary (candidate) containing a WD and a low-mass main-sequence (MS) star.}
Among the selected general periodic variable stars, a sample is found to be far deviated from the main sequence in CMD (Fig.~\ref{fig:periodic_flaring_CMD}), namely WFST~J044641.59+721910.2 
($\alpha=71.6733$ and $\delta=72.3195$, hereafter J0446+7219).
Its periodicity and sinusoidal-like light-curve profile are double checked by ZTF observations (see Fig.~\ref{fig:periodic_variable_J0446}).
J0446+7219 [$G_{\rm abs}=10.13$~mag and $(B_{\rm P}$-$R_{\rm P})_0=1.18$] locates in a CMD region corresponding to magnetic CVs \citep{lin+etal+2023+tmtsII} or WD-MS binaries \citep{Rebassa-Mansergas+etal+2021+WD_MS_binary}.
Due to the fact that magnetic CVs rarely exhibit sinusoidal-like light curves (see also \citealt{Liu+etal+2024+TMTS_CVs}),  we tend to support that J0446+7219 is an ellipsoidal binary with an orbital period of 5.90~hours, in which a low-mass MS is tidally deformed by its WD companion.
The spectroscopic identifications on Balmer lines and TiO bands could provide further evidences to support/refuse the inference.

\subsection{Short-period variable stars}

\begin{figure*}
    \includegraphics[width=0.96\textwidth]{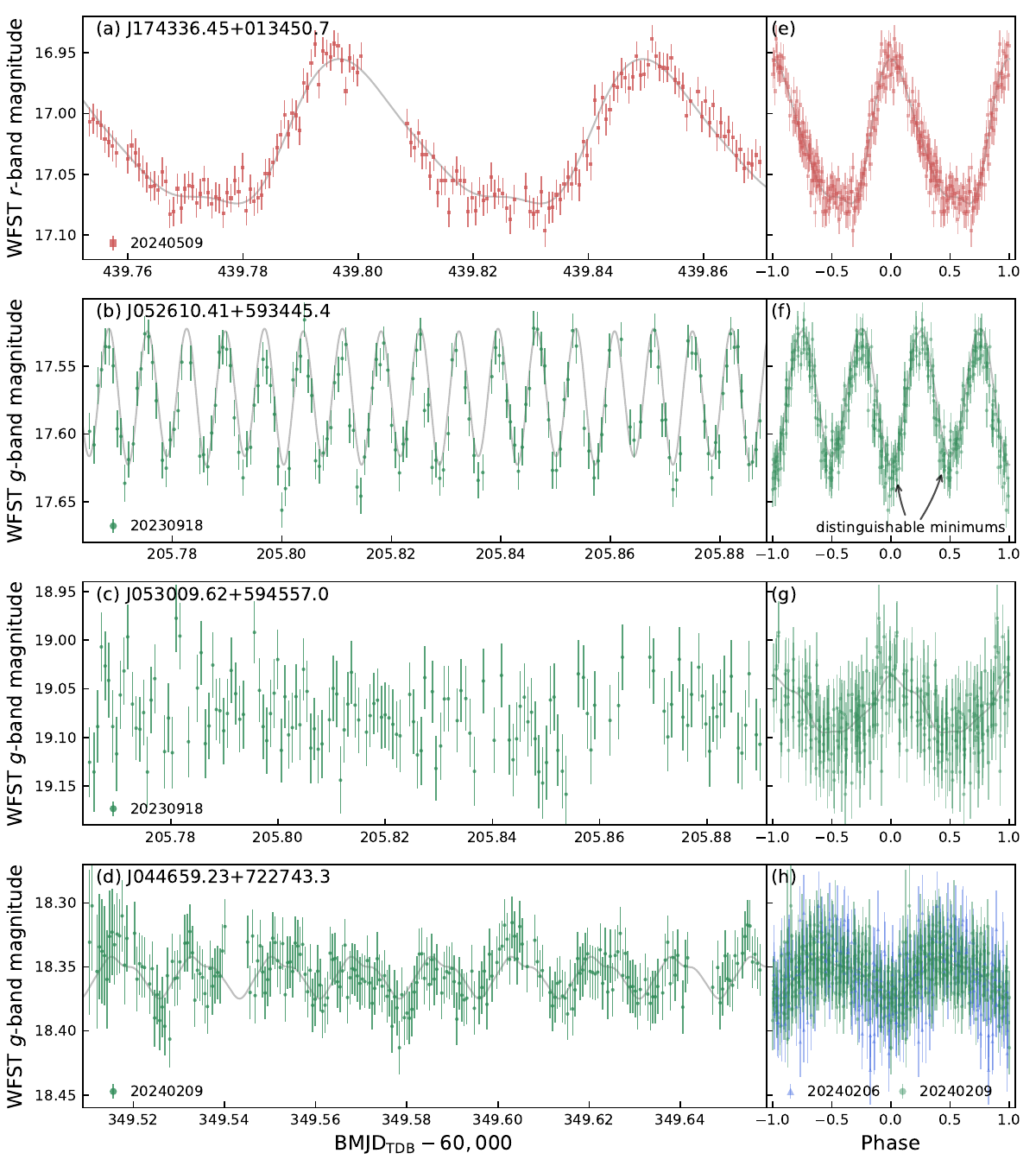}
    \caption{
    Uninterrupted light curves (panel a-d) and phase-folded light curves (panel e-h) for four short-period variable stars selected from WFST minute-cadence observations.
    These objects are 1.3-hour $\delta$ Scuti star J1743+0134 (panel~a \& e), 20.5-min ultracompact binary J0526+5934 (panel~b \& f), 6.7-min ZZ Ceti variable J0530+5945 (panel~c \& g), and 25.3-min unspecific variable hot WD J0530+5945 (panel~d \& h), respectively. To double check the periodicity, we present the phase-folded light curves for J0530+5945 obtained from two nights (panel~h).
    The grey solid lines represent the best-fit $4^{\mathrm{th}}$-order Fourier series.
     } 
    \label{fig:short_period_varibles}
\end{figure*}

\begin{figure*}
    \includegraphics[width=0.96\textwidth]{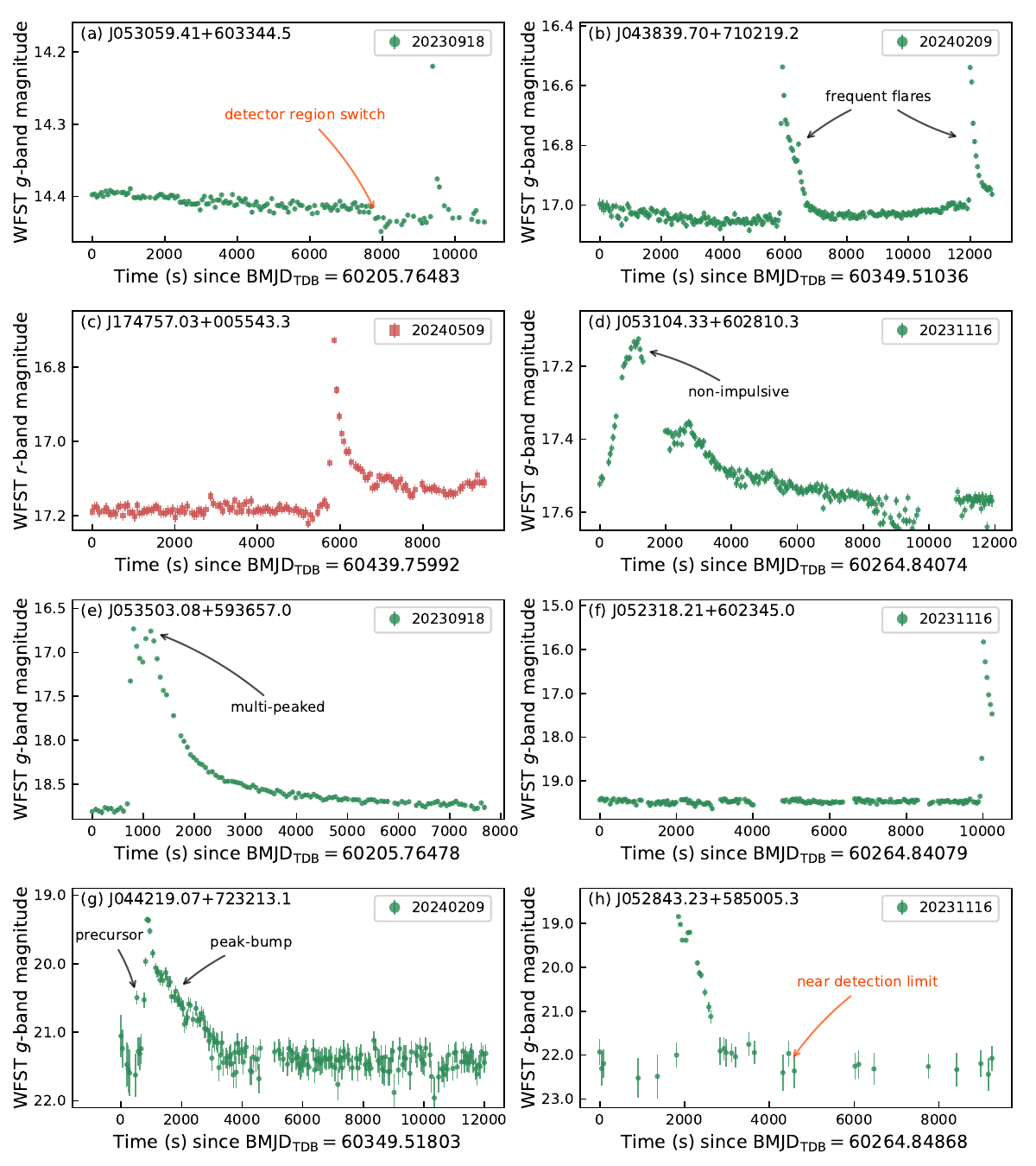}
    \caption{Uninterrupted light curves for eight flaring stars selected from WFST minute-cadence observations.
    All flaring stars are sorted by their quiescent magnitudes.
    We annotated light-curve morphology for the flares (black arrows) as well as the observation issues (red arrows).
    Notice that, since each ($4608 \times 4616$~pixel) region on the CCDs is processed separately in current pipelines, the bright sources switching among multiple detector regions could produce tiny systematic shifts in their light curves (see panel~a).
     } 
    \label{fig:flaring_star_varibles}
\end{figure*}

\begin{figure*}
    \includegraphics[width=0.96\textwidth]{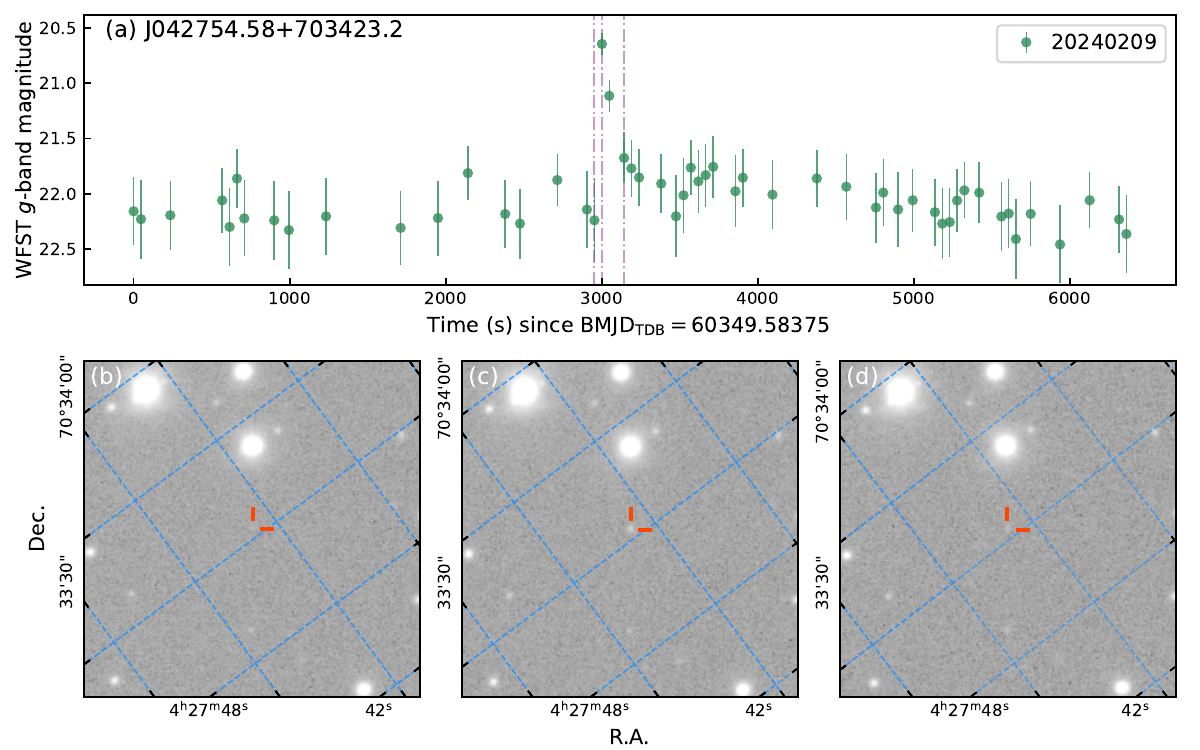}
    \caption{
    Uninterrupted light curve and finding charts for flaring star J0427+7034. The purple dotted-dashed lines overlapped on the light curve (panel~a) indicate the epochs corresponding to the finding charts (panel~b,c,d) in sequence.
    The red bars point the positions of J0427+7034 in the images. The size of each image is 300~pixels $\times$ 300~pixels.
     } 
    \label{fig:flaring_star_J0427}
\end{figure*}

As introduced in Section~\ref{sec:method_short_period}, 10 short-period variable stars were selected from the ${\rm Pwr}_{\rm max}$-$f_{\rm max}$ diagrams.
As the dominant ``noise'' sources in systematic search of interesting short-period variable stars \citep{lin+etal+2023+tmtsII}, hour-period $\delta$~Scuti stars are the main variable stars in our samples (e.g. the green stars within the classical instability strip of Fig.~\ref{fig:periodic_flaring_CMD}).
We present an example of $\delta$~Scuti star detected by WFST in panel~(a \& e) of Fig.~\ref{fig:short_period_varibles}.
With the increment in detection depths of survey telescopes, a large number of distant $\delta$~Scuti stars discovered by new instruments will be absence of credible parallax measurements from Gaia.
Hence, an identification tool that can distinguish $\delta$~Scuti stars from other short-period variable stars without the CMD, needs to be developed, e.g. the distribution of $\delta$~Scuti stars in period-amplitude diagram \citep{lin+etal+2023+tmtsII}.


To test WFST's capability in searching for and detecting short-period variable stars, 20.5-min ultracompact binary J0526+5934 \citep{Kosakowski+etal+2023+J0526_EW,lin+etal+2024+sdB_binary_NatAs,Rebassa-Mansergas+etal+2024+J0526_WD} was set as the targeted star of observation GP-20230918 and GP-20231116.
As panel~(b) of Fig.~\ref{fig:short_period_varibles} shows, despite a long dead time ($\approx 30$s) of camera, WFST can clearly reproduce the light curve for a 17.6-mag source having a variation period as short as 10.3 minutes (i.e. a half of its orbital period).
An outstanding feature is that its two minimums within an orbit are distinguishable in the WFST phase-folded light curve (panel~f of Fig.~\ref{fig:short_period_varibles}), which supports the asymmetric geometry between two ends of the tidally distorted visible component.
Such a feature from J0526+5934 has not been previously revealed by ZTF \citep{Kosakowski+etal+2023+J0526_EW}, Lijiang 2.4m Telescope \citep{lin+etal+2024+sdB_binary_NatAs} or 2.4m Thai National Telescope \citep{Rebassa-Mansergas+etal+2024+J0526_WD}. 

\textbf{One of the darkest ZZ~Ceti variables discovered yet.}
ZZ Ceti variables are a type of pulsating WDs with almost pure hydrogen atmospheres.
Thanks to the densely sampling data from Transiting Exoplanet Survey Satellite (TESS, \citealt{Ricker+etal+2015+TESS} ), dozens of the bright ($V<16$) ZZ Ceti variables have been discovered \citep{Romero+etal+2022+TESS_ZZceti,Romero+etal+2024+TESS_ZZ_more}.
However, due to rapid and relatively weak light variations, only a few ZZ Ceti variables darker than 19~mag have been identified yet \citep{Greiss+etal+2016+Kepler_ZZ}.
As introduced in Section~\ref{sec:method_short_period}, a significant 6.7-min periodic signal from J0530+5945 was revealed (see also panel~c \& g of Fig.~\ref{fig:short_period_varibles}).
Given the minute-level periodic source located exactly within the instability strip of ZZ Ceti (Fig.~\ref{fig:periodic_flaring_CMD}), we identified this object as a new member of ZZ Ceti variables.
Benefit from high detection sensitivity and accumulating (daily-cadence/minute-cadence) observational data,  a large number of ZZ Ceti variables will be discovered by WFST, offsetting the absence of dark pulsating WDs in the current survey missions.

\textbf{Unspecific variable hot white dwarf.} J044659.23+722743.3 ($\alpha=71.7468$ and $\delta=72.4620$, hereafter J0446+7227) is one of the bluest stars among the stars covered by our minute-cadence observations.
The color $(B_{\rm P}$-$R_{\rm P})_0=-0.58$ and absolute magnitude $G_{\rm abs}=10.89$~mag imply that J0446+7227 is a hot WD.
By interpolating the synthetic photometry tables provided from  MESA Isochrones \& Stellar Tracks (MIST, \citealt{Dotter+2016+MIST,Choi+etal+2016+MIST}), this color corresponds to an effective temperature as high as one hundred thousand Kelvin.
Within the observation GP-20240209, 25.3-min periodic variations from this hot WD were marginally revealed (see Section~\ref{sec:method_short_period}). 
Additionally, the phase-folded light curve obtained from the observation GP-20240206 agrees the periodic variations (panel~h of Fig.~\ref{fig:short_period_varibles} ).
Since its peak-to-peak amplitude is only 0.03~mag, which is comparable with WFST's photometric uncertainties at 18.3~mag, it is challenging for other survey missions to double check the periodicity.
As a hot WD with a (possible) variation period of about 25 minutes, J0446+7227 is inferred to be a candidate of variable DO WD (DOV, \citealt{Corsico+etal+2019+book+pulsatingWD} ) or rapidly rotating magnetic WD \citep{Ferrario+etal+1997+RoWD,Caiazzo+etal+2021+moon}.
Further spectroscopic and high-precision photometric observations could provide crucial clues to classify the object.

\subsection{Flaring stars}
\label{sec:flaring_stars}

Although dozens of time-domain survey missions are currently operating, only a few of them (e.g. TESS) are routinely performing observations in minute cadences, leading to the deficiency of stellar flares having densely observational sampling.
Based on the  minute-cadence observations of WFST, we present several well sampling examples of stellar flares, ranging from 14~mag to 22~mag, in Fig.~\ref{fig:flaring_star_varibles}.
In the figure, four ``energetic'' flares, J053104.33+602810.3 (panel~d),  J053503.08+593657.0 (panel~e), J044219.07+723213.1 (panel~g),  J052843.23+585005.3 (panel~h), were picked out by their high variabilities ($1/\eta$) rather than by the Osten's method.
Benefit from the densely sampling rates, these flaring stars exhibit clear light-curve profiles during flares, allowing us identify and describe temporal morphologies of flares \citep{Davenport+etal+2014+temporal_morphology,Howard+etal+2022+flare_morphology}, such as multiple peaks (panel~e of Fig.~\ref{fig:flaring_star_varibles}) and a highly impulsive peak following by a ``bump'' (i.e. peak-bump, panel~g).
The minute-cadence observations also revealed a short precursor emission prior to the large flare (panel~g), and a slow rising flare with a temporal structure deviated from the typical impulsive profile (panel~d).
Interestingly, a fast, large-amplitude flare was emitted from a very faint M dwarf J052843.23+585005.3 (panel~h). Its 22-mag quiescence magnitude is challenging to be detected by current time-domain survey missions, and thus such a dramatical flaring event could be incorrectly identified as an unknown fast optical transient (FOT).


\textbf{A minute-timescale flare emerged from a very dark WD binary candidate.}
J042754.58+703423.2 ($\alpha=66.9774$ and $\delta=70.5731$, hereafter J0427+7034) is a very faint object located below the red end of main sequence in the CMD (see Fig.~\ref{fig:periodic_flaring_CMD}).
Since its color $(B_{\rm P}$-$R_{\rm P})_0=2.91$ and absolute magnitude $G_{\rm abs}=13.51$~mag overlap on the evolutionary tracks of WD-MS binaries \citep{Rebassa-Mansergas+etal+2021+WD_MS_binary}, J0427+7034 is likely a binary consisting of a very dark WD and a M dwarf companion although the potential periodic light variations generated by its orbital motion are not detected.
From this object, a fast flare lasting for only a few minutes was occasionally captured by WFST (see panel~a of Fig.~\ref{fig:flaring_star_J0427}).
In order to confirm the abrupt light variation, we visually checked its images exposed before, on and after the flaring event:
the source was almost invisible before the event (panel~b), then it significantly became bright within a minute (panel~c) and decayed gradually (panel~d).
Such a transient flaring event from a 22-mag source is almost impossible to be captured by other survey instruments.
In the binary system of WD+M dwarf, the optical flares are likely induced by chromospheric activity of M dwarf or by accretion along the magnetic field lines of WD \citep{Pelisoli+etal+2023+WD_pulsar}.

\section{Summary} \label{sec:summary}
During the commissioning and pilot observation phases, we executed WFST minute-cadence observations on three different fields of Galactic plane, with a total on-source time of about 13 hours.
The accumulated fractions for the position deviations suggest that WFST's astrometric precision is better than 0.16~arcsec for 90\% astrometric measurements in these observations.
By cross-matching all photometric measurements within a radius of 1~arcsec, 650,696 uninterrupted light curves having at least 20~epochs, corresponding to 500,460 unrepeated WFST sources, were extracted from our observation data.

Via screening variable sources with variability index of $1/\eta$ and visually checking selected light curves,  we found that about 0.1\% of observed sources exhibit real astrophysical variability.
In order to select light curves covering fast flaring events, 486 flaring star candidates were selected from the uninterrupted light curves (with epochs $\geq$ 50) through the Osten's method, but only 33 ones show real flares by visually checking.
Additionally, 4 flaring stars exhibiting ``energetic'' flaring events were discovered from the variable star candidates with significantly high $1/\eta$ values.
By calculating the Lomb–Scargle periodograms for the uninterrupted light curves (with epochs $\geq$ 50) and plotting the ${\rm Pwr}_{\rm max}$-$f_{\rm max}$ diagrams, 10 short-period variable stars (with periods from several minutes to tens of minutes) were picked out. 
Additionally, 146 general periodic variable stars (with periods longer than 2~hours) were selected from splicing observations GP-20230918+GP-20231116 and GP-20240206+GP-20240209.

By cross-matching  500,460 WFST sources with Gaia DR3 catalog, we found that 97.87\% of them have at least a Gaia counterpart, and 20.93\% of these Gaia sources have a reliable parallax measurement, which support determining their positions in the CMD.
Referring to the catalog of WDs from Gaia EDR3 \citep{Gentile_Fusillo+etal+2021+WD_catalog}, there are at least 263 WD candidates covered by the WFST observations.

We presented the WFST light curves for a few periodic variable stars and flaring stars picked from our minute-cadence observations, and tried to identify the selected variable sources via their positions in the CMD (Fig.~\ref{fig:periodic_flaring_CMD}).
Besides the dominant EW-type eclipsing binaries and pulsating stars (e.g. RR Lyrae variables) among the selected general periodic variable stars, a 5.9-hr ellipsoidal binary candidate consist of a WD and a low-mass MS was first discovered from our observation data.
The selected short-period variable stars include a few $\delta$ Scuti stars, a known ultracompact binary (the targeted star of this field), a newly discovered and very faint ZZ Ceti variable, and an unspecific variable hot white dwarf.
Furthermore, our minute-cadence observations captured flaring stars from 14~mag to 22.3~mag, covering diverse flare temporal structures, such as multi-peaked and peak-bump flares.
We also highlighted a minute-timescale flare emerged from a candidate WD binary, while it is challenging for current survey missions to capture such a fast flare from a 22-mag star.

In conclusion, benefit from large aperture and thus deep detection limit, WFST can efficiently detect and reveal short-period variable stars and fast flaring stars in unexplored parameter spaces, thus leading to new opportunities in discovering unique variable sources in the northern sky.
Following the commencement of regular survey operations, we will perform a systematic analysis of WFST observation datasets and publish comprehensive catalogs for detected variable stars.

\begin{acknowledgments}
We acknowledge the anonymous referee for providing insightful comments.
The Wide Field Survey Telescope (WFST) is a joint facility of the University of Science and Technology of China, Purple Mountain Observatory.
This work is supported by National Key Research and Development Program of China (2023YFA1608100).
J.L. is supported by the National Natural Science Foundation of China (NSFC; Grant Numbers 12403038), the Fundamental Research Funds for the Central Universities (Grant Numbers WK2030000089), and the Cyrus Chun Ying Tang Foundations.
\end{acknowledgments}


\end{document}